\documentclass[aps,prd,twocolumn,showpacs,groupedaddress,nofootinbib]{revtex4}
\usepackage{graphicx}
\begin{document}

\title{Monte Carlo study of glueball masses in the Hamiltonian limit of
SU(3) lattice gauge theory}

\author{Mushtaq Loan}
\thanks{Email address: stsmushe@zsu.edu.cn}

\author{Xiang-Qian Luo}
\thanks{Corresponding author. Email address: stslxq@zsu.edu.cn}

\author{Zhi-Huan Luo}
\affiliation{Department of Physics, Zhongshan (Sun Yat-Sen) University,
Guangzhou 510275, China}

\date{\today}
\begin{abstract}
Using Standard Euclidean Monte Carlo techniques, we discuss in
detail the extraction of the glueball masses of 4-dimensional
SU(3) lattice gauge theory  in the Hamiltonian limit, where the
temporal lattice spacing is zero. By taking into account the
renormalization of both the anisotropy and the Euclidean
coupling, we calculate the string tension and masses of the
scalar, axial vector and tensor states using standard Wilson
action  on increasingly anisotropic lattices, and make an
extrapolation to the Hamiltonian limit. The results are compared
with estimates from various other Hamiltonian and Euclidean
studies. We find that more accurate determination of the glueball
masses and the mass ratios has been achieved and the results are a
significant improvement upon previous Hamiltonian estimates. The
continuum predictions are then found by extrapolation of results
obtained from smallest values of spatial lattice spacing. For the lightest
scalar, tensor and axial vector states we obtain masses of $m_{0^{++}}=1654
\pm 83$ MeV, $m_{2^{++}}=2272\pm 115$ MeV and $m_{1^{+-}}=2940\pm 165$ MeV,
respectively. 
These
are consistent with the estimates obtained in the previous studies in the Euclidean
limit. The consistency is a clear
evidence of universality between Euclidean and Hamiltonian
formulations. From the accuracy of our estimates, we conclude that
the standard Euclidean Monte Carlo method is a reliable technique
for obtaining results in the Hamiltonian version of the theory,
just as in Euclidean case.
\end{abstract}

 \maketitle

\section{INTRODUCTION}
In lattice QCD, Monte Carlo (MC) simulations on an Euclidean lattice is
the most popular method and is the preferred technique for
\emph{ab initio} calculations in QCD in the low energy regime.
 The calculation of the glueball spectrum is
of major importance as it may allow an interesting comparison of
QCD predictions derived from the first principles with the
experimental candidates for the glueball states such as the
$f_{0}(1710)$, $f_{2}(2400)$ or the $\eta (1440)$. 
 Whilst Euclidean lattice gauge theory (LGT) is expected to give the most
reliable estimates for QCD spectroscopy, the MC esults for glueball
masses have still been an issue under debate
\cite{Morningstar99,Morningstar97,Norman98,Teper98,Gupta91,Michael88,
Michael87,Bali93,Vaccarino99,Weingarten93,Lee99}. 
There are other areas where Euclidean
MC methods have been less successful. Examples are QCD at
finite temperature and density, glue
thermodynamics and heavy-quark spectra. 
Despite been rather neglected in
comparison to Euclidean LGT, the Hamiltonian
framework offers an interesting alternative
\cite{Luo:1996ha,Luo:96,Luo:1998dx,Gregory:1999pm,Luo:2000xi,Fang:2002rk,Luo:2004mc}
 that needs to be explored.
Hamiltonian formulation of lattice QCD has an appealing aspect in
reducing LGT to many-body problem such that
techniques familiar from quantum many-body theory and condensed
matter physics, such as the series expansions \cite{Hamer89}, the
$t$-expansion \cite{Horn84,Lana91} and the plaquette expansion
\cite{Hollenberg93,Hollenberg98} can be used to address the
problem. Another advantage of this formulation is that from a
numerical point of view, the reduction in the dimensionality of
the lattice from four to three dimensions  provides a significant
reduction in computational overheads.

Owing to the success of MC methods in the Euclidean
formulation, one might expect similar levels of success in
Hamiltonian regime. However MC approaches to the
Hamiltonian version of QCD have been less successful and lag at
least a decade behind the Euclidean calculations. While the
studies of the SU(N) glueball spectrum in 2+1 dimensions are
definitely feasible, the accurate studies of the SU(3) glueball
spectrum in 3+1 dimensions may be some way off in Hamiltonian
LGT. A number of quantum MC methods have
been applied to Hamiltonian LGT in the past, with somewhat mixed
results. One of the first attempts at such a calculation was
performed using a Green's function MC approach pioneered
by Hey and Stump for U(1) \cite{Heys83,Heys85} and SU(2)
\cite{Heys84} and later extended by Chin \emph{et al} for SU(3)
\cite{Chin86,Chin88,Chin88a}. However the later investigations
\cite{Hamer00} showed that this approach requires the use of a
``trial wave function" to guide random walkers in the ensemble
towards the preferred regions of configuration space. This
introduces a variational element into the procedure, in that the
results may exhibit a systematic dependence on the trial wave
function. A projector MC approach
\cite{Blankenbecler83,DeGrand85} and related ``stochastic
truncation" method \cite{Allton89} using a strong coupling
representation for the gauge field run into difficulties for
non-Abelian models, in that it requires Clebsch-Gordon
coefficients for SU(3) which are not even known at high orders.
The introduction of these coefficients cause destructive
interference between the transition amplitudes and a version of
the ``minus-sign problem" rears its head \cite{Hamer94}. In the
lack of any clear success of these methods, we are forced to
pursue an alternative approach.

In our previous studies, we used standard Euclidean MC
techniques to extract the Hamiltonian limit for the 3-dimensional
U(1) model \cite{Mushe03,Mushe03a,Mushe04} and  SU(3) lattice
gauge theory in 3+1 dimensions on anisotropic lattices
\cite{Mushe04a}. The idea is to measure observables on
increasingly anisotropic lattices and then extrapolate to the
Hamiltonian limit, corresponding to $\Delta\tau =
a_{t}/a_{s}\rightarrow 0$, where $a_{t}$ and $a_{s}$  are lattice
spacings in temporal and spatial directions. Applications of this
method to the above problems  
 have been extremely successful \cite{Mushe04,Mushe04a} and
have given rise to the great optimism about the possibility of
obtaining results relevant to continuum physics from MC
simulations of lattice version of the corresponding theory. The
objective of this work is to demonstrate the efficiency of this
method to determine the glueball masses in the Hamiltonian limit
of the SU(3) LGT on anisotropic lattices. In
numerical simulations on the anisotropic lattice one needs the
information upon the renormalization of anisotropy and couplings.
The importance of taking into account the difference of scales on
anisotropic lattices in these studies has already been discussed
by Klassen \cite{Klassen98}, through the renormalization of
anisotropy. In our case, the renormalization of Euclidean coupling
$\beta_{E}$ is also important in extracting our results. This may
be of some relevance to other studies on anisotropic lattices. We
find that the anisotropic Euclidean results converge to the
Hamiltonian estimates, once the difference of scales
\cite{Hasenfratz81} has been taken into account. Specifically we
find that this is particularly important at \emph{finite} anisotropy
\cite{Karsch82} in the extrapolation procedure.

The rest of the paper is organized as follows: In Sec. \ref{secII}, we
briefly discuss SU(3) gauge theory as defined on anisotropic
lattices. We also discuss our extrapolation
procedure to the Hamiltonian limit. The details of the
simulations, the methods used to extract the observables and the
analysis of the data, are described in Sec. \ref{secIII}. In this section
we also discuss our techniques for calculating static potential, string
tension and glueball masses from Wilson loop operators. The
determination of hadronic scale $r_{0}$ in terms of the lattice
spacing using the static potential is also outlined. 
We present and discuss our results in Sec. \ref{secIV}: the
glueball mass estimates are presented; finite-volume effects are
studied; results in the Hamiltonian limit are extracted.
extrapolations of our Hamiltonian estimates to the continuum limit
 and the conversion of our results into the physical units are
 made in Sec. \ref{secV}. Conclusions are given in Sec. \ref{secVI}.

\section{ANISOTROPIC DISCRETIZATION OF SU(3) THEORY}
\label{secII}
\subsection{Action}
The SU(3) gauge theory on the anisotropic lattice has the action \cite{Klassen98}
\begin{equation}
S_{g}[U] =
\frac{\beta_{s}}{\xi}\sum_{x}\sum_{i<j}\left[1-P_{ij}(x)\right] +
\beta_{\tau}\xi\sum_{x}\sum_{i}\left[1-P_{i4}(x)\right],
\label{eqn01}
\end{equation}
where the spatial and temporal plaquettes are given by
\begin{eqnarray}
P_{ij} & = & \frac{1}{3}\mbox{ReTr}\left[U_{i}(x)U_{j}(x+\hat{i})
U^{\dagger}_{i}(x+\hat{j})U^{\dagger}_{j}(x)\right],
\nonumber\\
P_{i4} & = & \frac{1}{3}\mbox{ReTr}\left[U_{i}(x)U_{4}(x+\hat{i})
U^{\dagger}_{i}(x+\hat{4})U^{\dagger}_{4}(x)\right],
 \label{eqn02}
\end{eqnarray}
with $x$ being the sites of the lattice, $i, j$ the spatial
directions, and $U_{i}(x)$  and $U_{4}$ the link variables in
spatial and temporal directions respectively. The
couplings\footnote{We have included different couplings in spatial
and temporal directions in order to allow the freedom to
renormalize so that correlation lengths are equal in both
directions, even though the lattice spacings $a_{s}$ and $a_{t}$
are different.} $\beta_{s}$ and $\beta_{\tau}$ are defined by
\begin{equation}
\beta_{s}= \frac{6}{g^{2}_{s}}, \hspace{1.0cm}
\beta_{\tau}=\frac{6}{g^{2}_{\tau}},
 \label{eqn03}
\end{equation}
and the anisotropy factor $\xi$ is defined by
\begin{equation}
\xi = \frac{1}{\Delta\tau} = \frac{a_{s}}{a_{t}}. \label{eqn04}
\end{equation}
Due to quantum fluctuations, the couplings and the anisotropy
deviate from their bare values. The $\xi$ dependence of the
couplings $g_{s}$ and $g_{\tau}$ leads to an energy sum rule for
the glueball mass, which differs in an important way from that
which one would expect naively. In the weak-coupling limit, the
relation between the scales of the couplings in Euclidean and
Hamiltonian formulations has been determined analytically from the
effective actions
\cite{Klassen98,Dashen81,Hasenfratz81,Hamer95,Karsch82}. Using the
background field technique of Dashen and Gross \cite{Dashen81},
Hasenfratz and Hasenfratz \cite{Hasenfratz81}, and Karsch
\cite{Karsch82} obtained a mapping between the equivalent
couplings of the Euclidean and anisotropic actions
\begin{eqnarray}
\frac{1}{g^{2}_{s}} & = &
\frac{1}{g^{2}_{E}}+c_{s}(\xi )+O(g^{2}_{E}),
\nonumber\\
\frac{1}{g^{2}_{\tau}} & = &
\frac{1}{g^{2}_{E}}+c_{\tau}(\xi)+O(g^{2}_{E}).
\label{eqn05}
\end{eqnarray}
For $\xi =1$ (Euclidean limit), one
recovers the Euclidean theory where $g_{s}=g_{\tau}=g_{E}$. In the
limit $\xi \rightarrow \infty$, Eqs. (\ref{eqn05}) reduce to the
Hamiltonian values as obtained in Refs.
\cite{Hasenfratz81,Hamer95}.

It is convenient to write Eq. (\ref{eqn01}) in a more symmetric way:
\begin{equation}
S_{g}[U] =
\frac{\beta_{\xi}}{\xi_{0}}\sum_{x}\sum_{i<j}\left[1-P_{ij}(x)\right]
+\beta_{\xi}\xi_{0}\sum_{x,i}\left[1-P_{it}(x)\right],
\label{eqn06}
\end{equation}
where
\begin{equation}
\beta_{\xi}=6/g^{2}_{\xi}, \hspace{1.0cm} \eta = \xi/\xi_{0}.
\label{eqn07}
\end{equation}
In the limit $\xi \rightarrow \infty$,
$\beta_{\xi}$ goes to the Hamiltonian coupling $\beta_{H}$. For
our calculations we use
\begin{eqnarray}
\beta_{\xi}& =& \beta_{E}+3\left[c_{s}(\xi )+c_{\tau}(\xi )\right]
+O(\beta^{-1}_{E}),
\nonumber\\
 \eta & = & 1+\frac{3}{\beta_{E}}\left[c_{s}(\xi )-c_{\tau}(\xi )\right]
+O(\beta^{-2}_{E}),
\label{eqn08}
\end{eqnarray}
where $\beta_{E}=6/g^{2}_{E}$. Therefore for every ($\beta_{E}
,\xi$) pair there is a corresponding pair of parameters
($\beta_{\xi} ,\xi_{0}$). For our simulations, we determine
$\beta_{\xi}$ by evaluating the factors $c_{s}(\xi )$ and
$c_{\tau}(\xi)$ by directly calculating them in terms of the
integrals given in Ref. \cite{Karsch82}. The values of
renormalization factor $\eta$ can be determined non-perturbatively
by matching Wilson loops in temporal and spatial directions\cite{Klassen98}. In our case this is however
only half the story, as there is also the renormalization of
coupling.

\subsection{Hamiltonian limit}
To obtain the Hamiltonian estimates we measure physical quantities
on increasingly anisotropic lattices, and make an extrapolation to
the extreme anisotropic limit $\xi\rightarrow \infty$. In the naive extrapolation
procedure, one might assume classical values of the couplings,
i.e., $\beta =\beta_{s}=\beta_{\tau}$ in Eq. (\ref{eqn01}) and
extrapolate the physical quantities to
the $\xi\rightarrow \infty$ limit at constant $\beta$. Such a procedure is not
strictly correct, however, at the quantum level because $\beta
\neq \beta_{s}\neq \beta_{\tau}$ due to renormalization
\cite{Karsch82,Klassen98}. The correct procedure is to extrapolate
the physical observables to $\xi \rightarrow \infty $ at constant
$\beta_{E}$. The procedure is summarized as
follows:

(i) Choose several anisotropies $\xi$ for a particular
$\beta_{E}$.

 (ii) Calculate the corresponding values of
$\beta_{\xi}$ and $\xi_{0}$, using Eq. (\ref{eqn08}).

 (iii) Use these couplings in Eq. (\ref{eqn06}) to do simulations and calculate physical
observables.

 (iv)Perform a polynomial fit to the data in the
inverse square anisotropy\footnote{This is because the leading discretization errors in
the temporal direction are expected to be of order $a_{t}^{2}$.}
$\Delta\tau^{2}$ at constant $\beta_{E}$ and extrapolate measured
quantities to $\Delta\tau \rightarrow 0$ ($\xi \rightarrow \infty
$).

As can be seen from Table \ref{tabpara}, the difference between bare and
renormalized couplings reaches a maximum around $\xi =4 (\Delta\tau
\approx 0.25)$, where the discrepancy
reaches nearly $\beta_{\xi}-\beta_{E} \approx 0.20$. Since the
most anisotropic lattices we use are in this vicinity significant discrepancies begin to creep in if the
correct procedure is not used.

\section{METHOD}
\label{secIII}

\subsection{Simulations parameters}
To evaluate string tension and glueball masses, we ran simulations
on four sets of lattices of different spatial extents. One such
set is listed{\footnote{It should be noted that Table
\protect\ref{tabpara} does not show all the used values of
parameters in this set of runs in the glueball measurements.} in
Table \ref{tabpara}. Configurations are generated using the mixture of
Cabibbo-Marinari (CM) \cite{Cabibbo82} pseudo-heat bath (where one
updates SU(2) subgroups of the SU(N) matrices) and over-relaxation
sweeps (which involves a large change in the link matrix to
increase the rate at which phase-space is traversed) with a ratio
\cite{Adler88,Brown87}. We define a {\emph compound sweep} as one
CM update followed by five over-relaxation sweeps. Configurations
are given a hot start and then 100 compound sweeps in order to
equilibrate. After thermalization, configurations are stored every
50 compound sweeps. We find that for physical observables, such as
static-quark potential and glueball masses, a suitable mixture of
heat bath and over-relaxation sweeps decorrelates field
configurations significantly, faster than a pure heat bath. The
successive configurations were found sufficiently independent to
justify the cost of evaluating glueball operators.

\begin{table}[!h]
\caption{ \label{tabpara} Run parameters used in the glueball
simulations.}
\begin{ruledtabular}
\begin{tabular}{cccccc}
Volume & $\beta_{E}$ & $\xi$ & $\beta_{\xi}$ &  $\xi_{0}$  \\  \hline
$8^{3}\times 12$ & 5.5  & 1.5  & 5.5943 & 1.4128   \\
$8^{3}\times 16$ & 5.5  & 2.0  & 5.6379 & 1.8261   \\
$8^{3}\times 20$ & 5.5  & 2.5  & 5.6986 & 2.2403   \\
$8^{3}\times 24$ & 5.5  & 3.0  & 5.6849 & 2.6683   \\
$8^{3}\times 32$ & 5.5  & 4.0  & 5.6881 & 3.5101   \\
$8^{3}\times 12$ & 5.6  & 1.5  & 5.6443 & 1.4143   \\
$8^{3}\times 16$ & 5.6  & 2.0  & 5.7495 & 1.8325   \\
$8^{3}\times 20$ & 5.6  & 2.5  & 5.7764 & 2.2525   \\
$8^{3}\times 24$ & 5.6  & 3.0  & 5.7850 & 2.6736   \\
$8^{3}\times 32$ & 5.6  & 4.0  & 5.7882 & 3.5179   \\
$8^{3}\times 12$ & 5.8  & 1.5  & 5.8943 & 1.4117   \\
$8^{3}\times 16$ & 5.8  & 2.0  & 5.9495 & 1.8378   \\
$8^{3}\times 20$ & 5.8  & 2.5  & 5.9764 & 2.2602   \\
$8^{3}\times 24$ & 5.8  & 3.0  & 5.9850 & 2.6837   \\
$8^{3}\times 32$ & 5.8  & 4.0  & 5.9882 & 3.5326   \\
$8^{3}\times 12$ & 6.0  & 1.5  & 6.0943 & 1.4197   \\
$8^{3}\times 16$ & 6.0  & 2.0  & 6.1495 & 1.8428   \\
$8^{3}\times 20$ & 6.0  & 2.5  & 6.1764 & 2.2675   \\
$8^{3}\times 24$ & 6.0  & 3.0  & 6.1850 & 2.6933   \\
$8^{3}\times 32$ & 6.0  & 4.0  & 6.1882 & 2.5464   \\
$8^{3}\times 12$ & 6.2  & 1.5  & 6.2943 & 1.4226   \\
$8^{3}\times 16$ & 6.2  & 2.0  & 6.3379 & 1.8442   \\
$8^{3}\times 20$ & 6.2  & 2.5  & 6.3986 & 2.2669   \\
$8^{3}\times 24$ & 6.2  & 3.0  & 6.3849 & 2.7020   \\
$8^{3}\times 32$ & 6.2  & 4.0  & 6.3881 & 3.5593   \\
$8^{3}\times 12$ & 6.4  & 1.5  & 6.4943 & 1.4245   \\
$8^{3}\times 16$ & 6.4  & 2.0  & 6.5379 & 1.8487   \\
$8^{3}\times 20$ & 6.4  & 2.5  & 6.5986 & 2.2735   \\
$8^{3}\times 24$ & 6.4  & 3.0  & 6.5849 & 2.7105   \\
$8^{3}\times 32$ & 6.4  & 4.0  & 6.5881 & 3.7167   \\
\end{tabular}
\end{ruledtabular}
\end{table}

\subsection{Static quark potential and string tension}
In order to convert the lattice observables into physical units,
one needs to set the scale by determining the lattice spacing for
each $\beta_{E}$ value, where the mass of the a low-lying particle is
typically used. Since the experimental
determinations of glueball masses are not clear, we use another
purely gluonic quantity, i.e., hadronic scale parameter $r_{0}$,
instead. This quantity can be measured very accurately on the
lattice\footnote{A disadvantage in using $r_{0}$ is that its
physical value must be deduced indirectly from the experiment and
there is some ambiguity in doing this.}. To determine $r_{0}$, in
terms of the lattice spacing $a_{s}$, we measure the static-quark
potential $V({\vec r})$ for various spatial separations ${\vec r}$
from the expectation values of the Wilson loops. The Wilson loops
are expected to behave like
\begin{equation}
W({\vec r},t) =
\sum_{l}A_{l}({\vec r})\mbox{exp}\left[-V_{l}({\vec r})t\right],
\label{eqn10a}
\end{equation}
where the summation is over the excited state contributions to the
expected value, and $l=1$ corresponds to the contribution from the
ground state. To obtain the optimal signal-to-noise ratio, it is
necessary to minimize the contamination from the excited states.
This is done by using APE smearing method \cite{Albanese87,Falcioni85}
which is
implemented by the iterative replacement of the original link by a
smeared link. Following the single-link procedure, a particular
spatial link is replaced by
 \begin{eqnarray}
 U_{i}(x) & \rightarrow &  P\left[(1-\alpha )U_{i}(x)+\frac{\alpha}{4}
 \sum_{j\neq i}\bigg(U_{j}(x)U_{i}(x+\hat{j})
 \right.
 \nonumber\\
 & & \left.
 \times U^{\dagger}_{j}(x+\hat{i})
 + U^{\dagger}_{j}(x)U_{i}(x-\hat{j})U_{i}(x-\hat{j})
 \right.
 \nonumber\\
 & & \left.
 \times
 U_{j}(x-\hat{j}+\hat{i})\bigg)\right] ,
 \label{eqn10b}
 \end{eqnarray}
 where the $P$ denotes a projection back onto SU(3) and $\alpha$
 is the smearing parameter. The parameter $\alpha$ is fixed to 0.7
 and six iterations of the smearing process are used. The optimum
 value of smearing is found by looking for a value that
 minimizes the statistical variation of the effective potential
 \begin{equation}
 V({\vec r},t) =
 -\mbox{ln}\left[\frac{W({\vec r},t)}{W({\vec r},t+1)}\right]
 \label{eqn10c}
 \end{equation}
 in the plateau region, while maintaining a good signal-to-noise
 ratio.
A plot of the static quark potential $V({\vec r})$ as a function of separation
$r$ at $\beta_{E}=5.8$ and $\Delta\tau = 0.333$ is shown in
Fig. \ref{figpoten}. The data in this plot were obtained by looking for the
 plateau in the effective potential.

The string tension is then extracted by fitting with the form
\begin{equation}
V({\vec r}) = V_{0}+Kr-\frac{e_c}{r},
 \label{eqn10d}
\end{equation}
where $V_{0}$, $K$ and $e_c$ are fit parameters. The string tension $\sigma$ is
found according to $K=a^{2}_{s} \sigma$. Only the on-axis
potential were used to extract the string tension. The errors in
the fit parameters were obtained using
bootstrap method. We see that the data are fitted very
well, in this instance giving $K = 0.199(4)$, $V_{0}= 0.682(9)$
and $e_{c}=0.162(1)$.

\begin{figure}[!h]
\scalebox{0.45}{\includegraphics{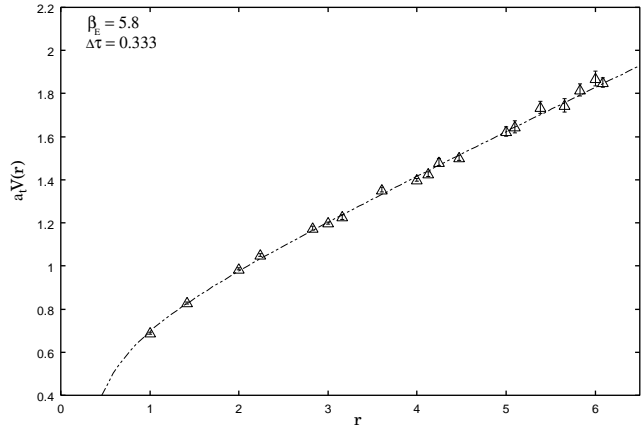}}
\caption{
\label{figpoten}
Static quark potential $V({\vec r})$ as a function of
separation $r$. The solid line is a fit Eq. (\ref{eqn10d}).}
\end{figure}

The hadronic scale $r_{0}$ is then determined using
\begin{equation}
\frac{r_{0}}{a_{s}} = \sqrt{\frac{(1.65-e_{c})}{\sigma a^{2}_{s}}} .
 \label{eqn10e}
\end{equation}
Table \ref{tab_hadronic} shows the results for $r_{0}/a_{s}$ along with the
results for the
string tension. Using these results, we can express the measured
quantities in terms of $r_{0}$.

\subsection{Glueball masses}
Our calculations are entirely conventional and follow essentially
the methods described in Ref. \cite{Morningstar97,Morningstar99}.
The glueball estimates are obtained from the time-like
correlations between space-like Wilson loops $\Phi (t)$,
\begin{equation}
C(t) = \langle \bar{\Phi}^{\dagger}(t)\bar{\Phi}(t_0)\rangle .
\label{eqn11}
\end{equation}
where 
\begin{equation}
\bar{\Phi}(t) = \Phi (t)-\langle 0\mid \Phi (t)\mid 0\rangle ,
\label{eqn12}
\end{equation}
 $\Phi$ is a gauge invariant operator
capable of creating a glueball out of vacuum. At large
$t$, and with periodic boundary conditions, the
correlator tends to behave as\footnote{
Due to the periodic boundary conditions, the correlation function
Eq. (\protect\ref{eqn13}), is a hyperbolic cosine function of $t-t_0$. If one assumes that
$T$ is much larger than $t-t_0$ at which we are able to evaluate the mass, one can
neglect the contribution from $exp[-m(T-(t-t_0))]$.}
\begin{eqnarray}
C(t) &= &Z \left[
\mbox{exp}\left(-m(t-t_0)\right) 
+ \mbox{exp}\left(-m \left(T-\left(t-t_0\right)\right) \right)\right]\nonumber\\
& &  + \mbox{excited state contributions},
\label{eqn13}
\end{eqnarray}
where $m$ is the effective mass of the lowest glueball state
which can be created by the operator $\bar{\Phi}(t)$. 
In the present study, we concentrate on
scalar, tensor and axial vector glueballs  by measuring the masses
in the representations
 $A_{1}^{++}$, $T_{2}^{++}$, $E^{++}$ and $T_{1}^{+-}$ respectively.

To determine mass $m$, $C(t)$  must be
calculated for large $t$ such that it is well
approximated by its asymptotic form. In principal we can use in
Eq. (\ref{eqn13}) any operator with the desired quantum numbers.
However, in a numerical calculation the statistical errors are
finite, and because $C(t)$ decreases
roughly exponentially in $t$, it will at large enough $t$, disappear
into the statistical noise. Thus the use of a glueball operator
for which $C(t)$ attains its asymptotic form as quickly as
possible is essential for extracting glueball mass. This requires
the need to use operators that are close to the wave-function of
the state in question.  To obtain the optimal signal-to-noise
ratio  we need large
smooth operators $\phi_{i}$ on the lattice scale. An optimized
operator is found by exploiting link smearing procedure
\cite{Falcioni85,Albanese87} and  variational techniques
\cite{Morningstar97,Morningstar99}. Such operators have  very good
overlap with the ground state.

Using a combination of single-link and double-link procedures,
glueball operators in
each of the $J^{PC}$ channels studied here,  are constructed from smeared
links $U^{s}_{j}(x)$ and fuzzy super-links $U^{s}_{j}(x)$.
A set of basic operators $\phi_{i}$ were constructed
using the linear combinations of the gauge-invariant path ordered
products of the smeared links about various space-like loops.  In
our simulations, these Wilson loops were measured from six
smearing schemes, each scheme
being a sequence of single-link and double-link mappings,
making 24  basic operators in all in
each of the channels investigated here. The optimized glueball
operator was then found as a linear combination of the basic
operators $\phi_{i}$
\begin{equation}
\Phi (t)=\sum_{i}v_{i}\phi_{i}(t),
\label{eqn16}
\end{equation}
where the coefficients $v_{i}$ were determined using the
variational technique. We computed  $24\times 24$ correlation
matrices\footnote{Note that the vacuum subtraction is required
only in the $A_{1}^{++}$ channel, since it has the identical
quantum numbers as that of the vacuum. The vacuum expectation
value vanishes identically in all other channels.}
\begin{eqnarray}
\tilde{C}^{(R)}_{ij}(t) & = & \sum_{t_{0} }\left[\langle
\phi^{(R)}_{i}(t+t_{0})\phi^{(R)}_{j}(t_{0})\rangle
\right.
\nonumber\\
& & \left.
- \langle
\phi^{(R)}_{i}(t+t_{0})\rangle\langle
\phi^{(R)}_{j}(t_{0})\rangle\right],
 \label{eqn17}
\end{eqnarray}
for $R=A_{1}^{++}$, $T_{2}^{++}$ (or $E^{++}$), and $T_{1}^{+-}$ which correspond to
$0^{++}, 2^{++}$ and $1^{+-}$ states respectively. We expect that
\begin{eqnarray}
\tilde{C}^{(R)}_{jk}(t) & = & \langle \phi^{(R)}_{j}\mid \mbox{e}^{-Ht}\mid
\phi^{(R)}_{k}\rangle \nonumber\\
& & = \sum_{i}\langle \phi_{j}^{(R)}\mid\phi_{i}\rangle \langle
\phi_{i}\mid\phi^{(R)}_{k}\rangle \mbox{e}^{-m_i^{(R)}t},
\label{eqn18}
\end{eqnarray}
where $\mid\phi_{i}\rangle$ are the eigenstates of the
Hamiltonian. At large temporal separation, the lowest mass
$m^{(R)}$ dominates; it belongs to a glueball state which in
the continuum limit will have the lowest spin contained in the
representation $R$.

The  coefficients $v^{(R)}_{i}$ in Eq. (\ref{eqn17}) were
determined by minimizing the effective mass
\begin{equation}
\tilde{m}^{(R)}(t) =
-{1 \over t-t_0} \mbox{ln}\left[\frac{\sum_{ij}v^{(R)}_{i}
v^{(R)}_{j}\tilde{C}^{(R)}_{ij}(t)}{
\sum_{ij}v^{(R)}_{i}v^{(R)}_{j}\tilde{C}^{(R)}_{ij}(t_0)}\right].
\label{eqn19}
\end{equation}

\subsection{Analysis details}
In the analysis phase, the glueball masses were extracted using a
two-step procedure. First we solved the generalized eigenvalue
problem
\begin{equation}
\tilde{C}(t){\bf{v}}^{(R)} =\omega (t)\tilde{C}(t_0){\bf{v}}^{(R)}, 
\label{eqn21}
\end{equation}
derived from minimization of Eq. (\ref{eqn19}). 
The correlation matrices $\tilde{C}$ are then
projected to the space of eigenvectors corresponding to the $N$
maximum eigenvalues with which  satisfy the condition
\begin{equation}
\omega_{i} (t) > \varepsilon , ~~ i=1, \cdots, N,
 \label{eqn22}
\end{equation}
where $\varepsilon$ is a small adjustable parameter. The reduced
correlator\footnote{The index $R$ for the
representation is now omitted.} is
\begin{equation}
C^{N} (t) =  {\bf{v}^T}\tilde{C}^N(t){\bf{v}}  .
\hspace{0.50cm} 
\label{eqn23}
\end{equation}
The parameter $\varepsilon$ is chosen appropriately so that we get
rid of the unphysical modes caused by negative and very small
eigenvalues\footnote{We chose $\varepsilon$ in Eq.
(\protect\ref{eqn22}) so that the reduced matrix had dimension
between 2 and 10. The result then does not depend on $\varepsilon$.}. 
We read the effective glueball mass directly from
the largest eigenvalue corresponding to the lowest energy
\begin{equation}
m_{eff}=-\mbox{ln}\left[\frac{\omega_{0}(t+1)}
{\omega_{0}(t)}\right].
\label{eqn24}
\end{equation}
Alternatively, one can obtain the effective mass from
\begin{equation}
m_{eff}=-\mbox{ln}\left[\frac{C^N(t+1)}{C^N(t)}\right] ,
\label{eqn25}
\end{equation}
by setting $t-t_0=1$ in Eq. (\ref{eqn23}), as Refs. \cite{Morningstar99,Morningstar97}.

Fig. \ref{figeffmassplot} graphs an effective mass showing the results
of a single
exponential to the optimized correlation function for the
$A_{1}^{++}, T_{1}^{+-}$ and $T_{2}^{++}$ channels for $\beta_E
=5.8$ and $\Delta\tau =0.5$. The signal is seen out to the time slice 7
and reaches a plateau region for $1\leq t \leq 7$ in each channel.
The data are little noisy for $t > 7$. It can be seen that  two different
evaluations of Eqs. (\ref{eqn24}) and (\ref{eqn25}) yield very consistent results
within statistical errors.

\begin{figure}[!h]
\scalebox{0.45}{\includegraphics{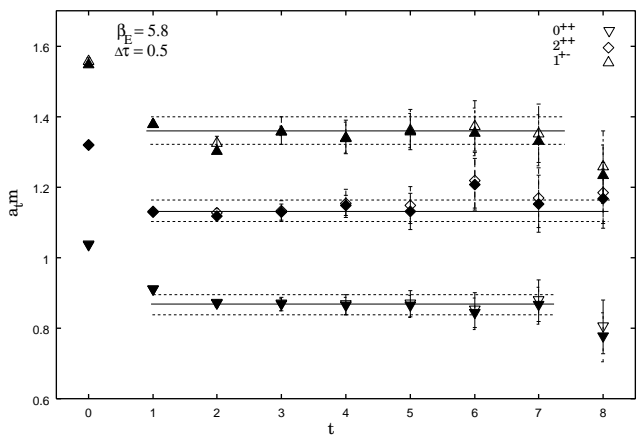}}
\caption{
\label{figeffmassplot}
Effective mass plot for $A_{1}^{++},T_{2}^{++}$.
The open and filled symbols correspond to Eqs. (\protect\ref{eqn24})
and (\protect\ref{eqn25}) respectively.}
\end{figure}

\section{SIMULATION RESULTS}
\label{secIV}

Here we focus on the lightest states because the correlations mediated
by heavier states decrease so rapidly with $t$ that it becomes
hard to know whether we have indeed isolated the asymptotic
exponential decay. Moreover, glueballs that are heavy enough will
decay into lighter glueballs and this may require more careful
analysis. The   variationally
optimized glueball correllators were calculated on the all
lattices given in Table \ref{tabpara}. For each glueball channel we
find a region $t \in [t_{min}, t_{max}]$ in which the correlation
function was well described by its asymptotic form Eq. (\ref{eqn13}). 
To extract the
masses from our data sets, we examine the onset of the plateaus in
all effective masses. In most cases, acceptable fit results were
obtained from the fit interval $t \in [2,6]$.

Fig. \ref{figeffA1ppplot} shows the effective scalar mass
and scalar mass fit as a function of  $t$ 
for $\beta_{E}$=6.0 and 6.2 at $\Delta\tau=0.333$. As a result of the
link-smearing and variational techniques, a significant signal is found
and an impressive plateau is observed in the effective mass plot.

\begin{figure}[!h]
\scalebox{0.45}{\includegraphics{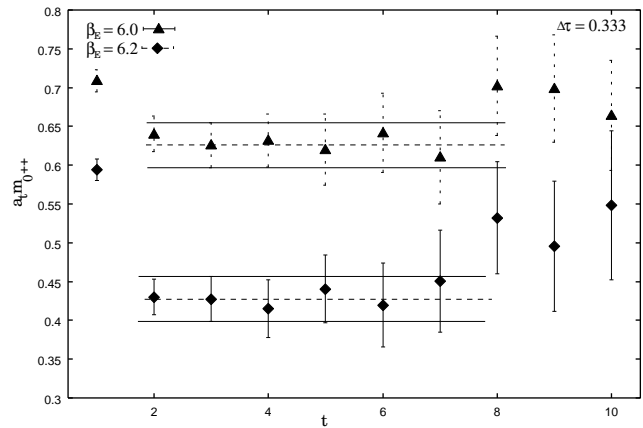}}
\caption{
\label{figeffA1ppplot}
Effective mass plot for $A_{1}^{++}$  channel at
$\beta_{E}=6.0$ (solid triangles) and $6.2$ (solid diamond) for
$\Delta\tau =0.333$.}
\end{figure}

Fig. \ref{figeffT2T1} shows the effective mass plot for tensor and
axial vector glueballs for  $\beta_{E} = 6.0$ and $\Delta\tau =
0.333$. The data at lower $\beta_{E}$ values ($\beta_{E} < 5.4$)
turned out to be too noisy to allow for reliable estimates of the
masses, with no clear plateau visible and errors much larger than
the data. A possible reason for this faster exponential drop-off
of the correlation functions is  that  at the strong coupling
(smaller $\beta_{E}$, and so larger $a$), the statistical errors
overwhelm the signal at smaller values of $t$ and
no useful effective mass beyond $t=2$ is obtained. In these cases
one simply assumes that $m_{eff}(t=2)$ provides a good mass
estimate. The effective mass plateau for the heavier states
becomes less significant due to  very large statistical errors at
smaller $t$ values. We have not included our estimates for tensor
and axial vector glueballs for  $\beta_{E} < 5.4$ in our final
analysis of mass gaps and mass ratios. The signal improves and the
plateau becomes more visible as we walk towards the weak-coupling
region. One, however,  does expect large finite-size correction as
one moves towards the weak coupling region. We comment on
this below. We measure both the
$E^{++}$ and $T_{2}^{++}$ to check the expected degeneracy and the
restoration of rational invariance, since these states contribute
to $J^{PC}=2^{++}$ in the continuum $O(3)$ symmetry group.

\begin{figure}[!h]
\scalebox{0.45}{\includegraphics{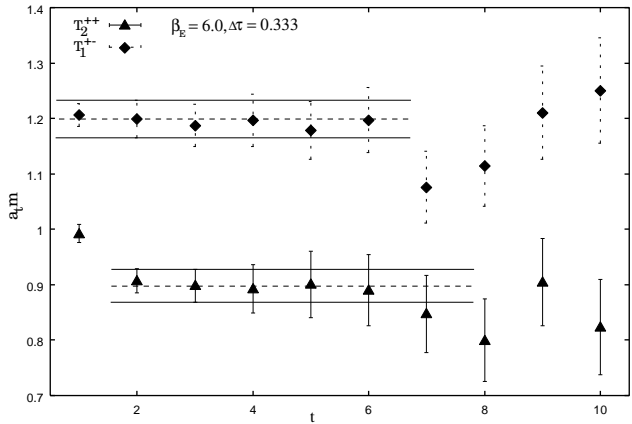}}
\caption{
\label{figeffT2T1}
Effective mass plot showing the results of a single exponential fit
of glueball correlation functions for $T_{2}^{++}$ and
$T_{1}^{+-}$ channels for $\beta_{E}=6.0$ and $\Delta\tau =0.333$.}
\end{figure}

Our  results for the glueball masses in the lattice units for
various $\beta_{E}$ at $\Delta\tau=0.50$ and $0.25$ are shown in
Tables \ref{tab_SM1} and \ref{tab_SM2} respectively. All errors
were obtained by jackknife method.

\begin{table}[!h]
\caption{ \label{tab_hadronic} Results for the string tension
$K=a_{s}^{2}\sigma$, hadronic scale parameter $r_{0}$ in terms of
lattice spacing $a_{s}$ and the Coulomb coupling $e_{c}$ from
$\Delta\tau = 0.50$ simulations.}
\begin{ruledtabular}
\begin{tabular}{cccc}
$\beta_{E}$ &  $e_{c}$ &  $K (=a_s^{2}\sigma)$ & $a_{s}/r_{0}$
\\\hline
5.0  & 0.3611(8) & 0.81(5)   &0.79(4) \\
5.2  & 0.341(2)  & 0.73(2)   &0.75(1) \\
5.4  & 0.4127(6) & 0.601(7)  &0.699(7) \\
5.5  & 0.329(3)  & 0.431(4)  &0.571(6) \\
5.6  & 0.296(2)  & 0.273(2)  &0.449(4) \\
5.8  & 0.283(2)  & 0.1103(8) &0.284(2) \\
6.0  & 0.265(1)  & 0.0683(5) &0.222(1) \\
6.2  & 0.302(2)  & 0.0364(2) &0.164(1) \\
6.4  & 0.298(2)  & 0.0227(1) &0.129(1) \\
\end{tabular}
\end{ruledtabular}
\end{table}

\begin{table}[!h]
\caption{ \label{tab_SM1} Glueball masses in the lattice units
from $\Delta\tau = 0.50$ simulations.}
\begin{ruledtabular}
\begin{tabular}{cccccc}
$\beta_{E}$  & $a_{t}m_{A_{1}^{++}}$ & $a_{t}m_{T_{2}^{++}}$ &
$a_{t}m_{E^{++}}$ & $a_{t}m_{T_{1}^{+-}}$
\\\hline
5.0   & 2.12(3)  &  &   &  & \\
5.2   & 1.74(4)  &  &   &  & \\
5.4   & 1.43(2)  &2.11(2) & 2.06(3) & 2.48(4)\\
5.5   & 1.274(8) &1.68(3) &1.62(4)  & 1.93(3)\\
5.6   & 1.133(6) &1.51(2) &1.48(3)  & 1.72(2)\\
5.8   & 0.87(2)  &1.11(2) &1.075(8) & 1.35(2)\\
6.0   & 0.663(6) &0.890(8)&0.85(4)  & 1.154(8)\\
6.2   & 0.564(4) &0.721(3)&0.684(4) & 0.942(5)\\
6.4   & 0.433(3) &0.647(4)&0.614(6) & 0.730(2)\\
\end{tabular}
\end{ruledtabular}
\end{table}

\begin{table}[!h]
\caption{ \label{tab_SM2} Glueball masses in the lattice
units from $\Delta\tau = 0.25$ simulations.}
\begin{ruledtabular}
\begin{tabular}{ccccccc}
$\beta_{E}$ & $a_{t}m_{A_{1}^{++}}$ & $a_{t}m_{T_{2}^{++}}$ &
$a_{t}m_{E^{++}}$ & $a_{t}m_{T_{1}^{+-}}$
\\\hline
5.0  &  2.01(1)  &  &   &  & \\
5.2  &  1.64(2)  &  &   &  & \\
5.4  &  1.427(8) &2.10(1) & 1.99(3) & 2.37(2)\\
5.5  &  1.182(5) &1.603(9)&1.536(7) &1.92(1)\\
5.6  &  1.027(3) &1.37(1) &1.26(2)  &1.56(1)\\
5.8  &  0.81(1)  &1.08(1) & 1.02(2) &1.31(2)\\
6.0  &  0.598(3) &0.841(3)& 0.80(4) &1.116(6)\\
6.2  &  0.511(1) &0.690(3)& 0.624(4)&0.892(3)\\
6.4  &  0.410(1) &0.563(3)& 0.514(4)&0.727(3)\\
\end{tabular}
\end{ruledtabular}
\end{table}

Among the lattice artifacts,  the finite volume effects pose a
major problem in the glueball mass determinations from coarse
lattice simulations. The mass of particles confined in a box with
periodic boundary conditions can differ appreciably from their
infinite-volume values. It is being suggested that finite-size
effects can also induce a mass splitting of the $E$ and $T$
polarizations.

To measure the systematic errors in our estimates from finite
volume, we calculate  masses on volumes $a_s L_{1}$ and $a_s L_{2}$
which are both significantly larger than hadronic length scale and
measure the changes in the glueball masses as the volume is
changed. Essentially it tells us that if the change in masses is
small when we go from $L_{1}$ to $L_{2}$, then we can be confident that
masses calculated on the largest volume are identical, within
statistical errors, to the infinite volume limit masses. This
strategy to control finite volume corrections, however, requires
large simulations over a number of $\beta_{E}$ values on a range of
lattice volumes. In practice one chooses a couple of $\beta_{E}$
values to perform analysis on large volumes. This allows one to
establish what volumes are large enough that any mass shift
becomes visible within statistical errors.

In order to ascertain the volume effect on our glueball
measurements, we performed extra
simulations at $\beta_{E} = 6.0,6.2$ and $6.4$ over a range of
anisotropies, $\Delta\tau= 0.666-0.25$ (since we also
intend to examine the Hamiltonian limit at these couplings).
The spatial length varies from $L= 10$ to $L=16$ which
corresponds to a
variation from 0.90 - 1.5 fm, a range of sizes that satisfies the naive
condition that spatial extent should be around or
greater than $2r_0$.
The input parameters used in these runs are same as those used in the
$8^{3}\times T$ volume simulations. The results
from the $L= 10, 12$ and 16 runs for the glueball masses
in lattice units
are given in Tables \ref{tab_lvol1}, \ref{tab_lvol2} and \ref{tab_lvol3}.
We plot the masses against the spatial size in
Fig. \ref{figfinitevol}.

\begin{figure}[!h]
\scalebox{0.45}{\includegraphics{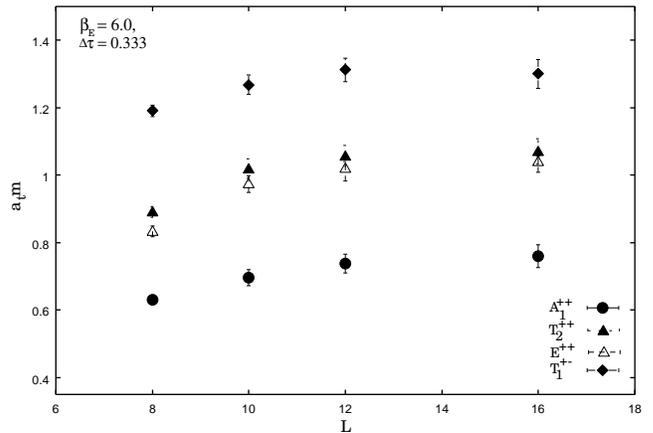}}
\caption{
\label{figfinitevol}
Variation of effective glueball masses with the spatial size $L$.}
\end{figure}

\begin{table}[!h]
\caption{ \label{tab_lvol1} The glueball masses in the
lattice units for various spatial extensions from
$\beta_{E} = 6.0 $ and $\Delta\tau = 0.333$ simulations.}
\begin{ruledtabular}
\begin{tabular}{cccccc}
$L$ & $a_{t}m_{A_{1}^{++}}$ & $a_{t}m_{T_{2}^{++}}$ & $a_{t}m_{E^{++}}$
& $a_{t}m_{T_{1}^{+-}}$ \\\hline
8   & 0.63(1) & 0.89(2) &0.83(2) &1.19(2) \\
10  & 0.69(2) & 0.99(3) &0.94(3) &1.26(3)  \\
12  & 0.73(2) & 1.05(3) &1.00(3) &1.31(3)  \\
16  & 0.76(3) & 1.06(4) &1.04(4) &1.30(4) \\
\end{tabular}
\end{ruledtabular}
\end{table}

There are few observations we can make from Fig. \ref{figfinitevol}
and Tables
\ref{tab_lvol1}, \ref{tab_lvol2} and \ref{tab_lvol3}.

(i) We do indeed observe substantial finite size effects, as our results
at $\beta_E =$ 6.0, 6.2 and 6.4
were found to differ appreciably from those  obtained on larger
 volumes.

(ii) The $T_{2}^{++}$ and $E^{++}$ states show a
separation at small volume, indicating the breaking of
rotational symmetry. This  separation, however, decreases as $L$
increases. This has been observed for tensor glueball mass for
SU(2) \cite{Michael88}. The reason for this  change
lies in the possible contamination of the small lattice glueball
configurations by those due to torelon states (flux excitations
encircling the toroidal lattice) \cite{Michael87}. Thus $E^{++}$
state seems to behave more like a torelon state for small spatial
size.

(iii) We infer from Fig. \ref{figfinitevol} that the results from larger
volume  differ very little. We find that for  $\beta_{E} \geq 6.0$  a
spatial size of $L \geq 12$ is large enough for the lightest glueballs
 to be free of finite size effects within our statistical errors.

\begin{table}[!h]
\caption{ \label{tab_lvol2} Glueball masses in the
lattice units for various spatial extensions from
$\beta_E= 6.2 $ and $\Delta\tau = 0.25$ simulations.}
\begin{ruledtabular}
\begin{tabular}{cccccc}
$L$ & $a_{t}m_{A_{1}^{++}}$ & $a_{t}m_{T_{2}^{++}}$ & $a_{t}m_{E^{++}}$
& $a_{t}m_{T_{1}^{+-}}$ \\\hline
8   & 0.511(1) & 0.690(3) &0.62(2)  &0.892(3) \\
10  & 0.570(6) & 0.747(7) &0.709(5) &0.974(8) \\
12  & 0.598(8) & 0.783(9) &0.768(6) &1.05(1) \\
16  & 0.582(1) & 0.79(1)  &0.78(2)  &1.06(1) \\
\end{tabular}
\end{ruledtabular}
\end{table}

\begin{table}[!h]
\caption{ \label{tab_lvol3} Glueball masses in the
lattice units for various spatial extensions from
$\beta_{E} = 6.4 $ and $\Delta\tau = 0.333$ simulations.}
\begin{ruledtabular}
\begin{tabular}{cccccc}
$L$ & $a_{t}m_{A_{1}^{++}}$ & $a_{t}m_{T_{2}^{++}}$ & $a_{t}m_{E^{++}}$
& $a_{t}m_{T_{1}^{+-}}$ \\\hline
8   & 0.418(1) & 0.597(2) &0.567(2) &0.716(4) \\
10  & 0.470(4) & 0.647(5) &0.636(5) &0.774(6) \\
12  & 0.462(3) & 0.653(3) &0.648(6) &0.789(5) \\
16  & 0.467(3) & 0.643(2) &0.641(4) &0.775(4) \\
\end{tabular}
\end{ruledtabular}
\end{table}

For sufficiently large $L$, the scalar and tensor glueball masses
$m_{0}(L)$ and $m_{2}(L)$ deviate from
their infinite volume limits, $m_{0}(\infty )$ and $m_{2}(\infty)$,
respectively, by \cite{Luscher86}
\begin{eqnarray}
a_{t}m_{0,2}(L)& = & a_{t}m_{0,2}(\infty
)\left[1-\rho_{G}\frac{\mbox{exp}\big(-\sqrt{3}m_{0}L/2\big)}{m_{0}L}
\right.
\nonumber\\
& & \left. - O\bigg(\frac{\mbox{exp}(-m_{0}L)}{m_{0}L}\bigg)\right],
\label{eqn26}
\end{eqnarray}
where $\rho_{G}$ is related to the strength of an
effective triple scalar glueball interaction vertex.  The mass
shift in above equation arises from the exchange of scalar
glueballs across the boundaries of the lattice.

Fitting the data at larger volumes to the first two leading terms of Eq.
(\ref{eqn26})
yields the glueball mass in the infinite volume limit. We did the $A_{1}^{++}$
fit first using the above fit function with $a_{t}m_{0}(\infty)$ and $\rho_{G}$ 
as fitting
parameters. A fit to the scalar data at
$\beta_{E} =6.0$ and $\Delta\tau = 0.25$
 yields $a_{t}m_{0^{++}}(\infty)= 0.735\pm 0.011$  and $\rho_{0}=184\pm 27 $.
The best-fit values for $a_{t}m_{0^{++}}(\infty)$ was then used in the fits to
the results  for other two channels. A similar fit the tensor data
gives $a_{t}m_{2^{++}}= 1.049\pm 0.022$ and $\rho_{2}=144\pm 19$. 
We also used Eq. (\ref{eqn26}) for the axial vector. Overall,
it appears that at $\beta_{E} =6.0$, $a_{t}m_{0,1,2}(8)- a_{t}m_{0,1,2}(\infty)$ is of
the order of $10 - 15\%$. The results of these fits are summarized
in Table \ref{tab_SM3}.

\begin{table}[!h]
\caption{ \label{tab_SM3} The glueball estimates in
the infinite volume limit for $\Delta\tau =0.25$ at various
$\beta_{E}$ values.}
\begin{ruledtabular}
\begin{tabular}{ccccc}
$\beta_{E}$ & $a_{t}m_{A_{1}^{++}}(\infty )$ & $a_{t}m_{T_{2}^{++}}(\infty )$
& $a_{t}m_{T_{1}^{+-}}(\infty )$ \\\hline
6.0   & 0.73(1) & 1.04(2)    &1.33(2) \\
6.2   & 0.58(1) & 0.78(3)    &1.06(4) \\
6.4   & 0.483(5)& 0.637(8)   &0.779(8) \\
\end{tabular}
\end{ruledtabular}
\end{table}

The mass estimates in $L \to \infty$ at various
anisotropies are then extrapolated to the Hamiltonian limit,
$\Delta\tau \rightarrow 0$. To maintain the accuracy of the final results
we choose all anisotropic points on the larger volumes and make an
extensive analysis. This allows us to stay  close to the Hamiltonian limit
and translate to $\Delta\tau \rightarrow 0$. With the extrapolation procedure
used here, the results are almost as accurate at large anisotropies
as in the isotropic and the extrapolation can be made with confidence.

\section{FINAL RESULTS FOR STRING TENSION AND GLUEBALL MASSES}
\label{secV}

\subsection{Extrapolation to the Hamiltonian limit}

The extrapolation of the string tension to the Hamiltonian limit,
performed in the powers of $\Delta\tau^{2}$,  is shown in
Fig. \ref{figextstring}. Unlike our previous estimates \cite{Mushe04a}
where we observed a trend towards a strong curvature in the extrapolation
for the intermediate values of $\beta_{E}$, we see a fairly smooth variation
of the string tension with $\Delta\tau^{2}$. In the weak-coupling region
$K$ remains rather independent of $\Delta\tau^{2}$, thus
enabling reliable extrapolation to the Hamiltonian limit.
\begin{figure}[!h]
\scalebox{0.45}{\includegraphics{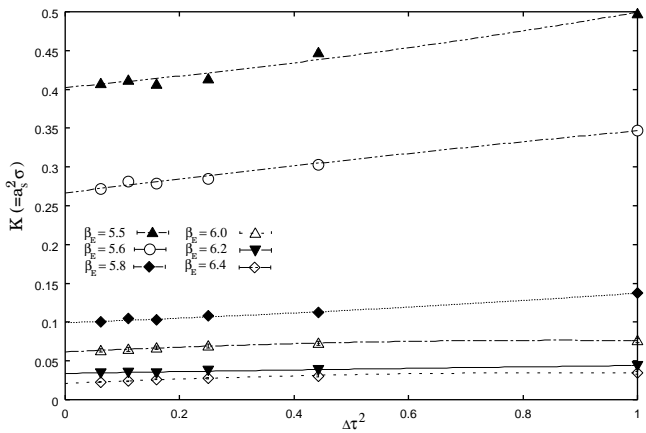}}
\caption{
\label{figextstring}
Extrapolation of the string tension to the Hamiltonian limit $\Delta\tau
\rightarrow 0$. The curves are the quadratic fits in $\Delta\tau^{2}$.}
\end{figure}

To check the accuracy of our method, we compare our extrapolated
estimates with Hamiltonian estimates obtained by using an exact linked
cluster expansion (ELCE) \cite{Hamer86}.
However, for such a comparison we must
take into account the difference of scales between Euclidean and
Hamiltonian regimes. The Hamiltonian coupling parameter
$\lambda=6/g^{4}_{H}$, where
$g_{H}=\lim_{\xi\rightarrow\infty}g_{\xi}$ may be related to 
$\beta_{E}$ through the relation \cite{Hasenfratz81}
\begin{equation}
\beta_{E} =\sqrt{6\lambda} - 0.07848.
\label{eqn30a}
\end{equation}

 We see that  our results
agree well with those obtained from ELCE although less accurate in the
small $\beta_{E}$ region. We do not expect precise agreement in this
 region, since ELEC results correspond to the energy per unit length of
a ``string'' of flux along one axis, whereas our estimates refer to the
decay exponent
of space-like Wilson loops\footnote{It is well know that
at the strong coupling, where rotational invariance is broken,
these two estimates differ at strong coupling.
However beyond roughening transition \protect\cite{Hasenfratz3,Lusher81},
where the
rotational symmetry is restored, different estimates of the string tension
should coincide.}.
We do not show here the GFMC results \cite{Hamer00}, which were obtained from
Creutz ratios on very small loops, and therefore subjected to large finite-size
effects. Our results in the Hamiltonian limit for the string tension $K$,
 hadronic scale $r_{0}$ in terms of the lattice spacing $a_{s}$
 and the Coulomb coupling $e_c$  
are shown in Table \ref{tab_HSTresults}. Also are shown, for
comparison, the previous Hamiltonian estimates from ELCE.

\begin{table}[!h]
\caption{ \label{tab_HSTresults} Results for the string tension $K$ and
hadronic scale $r_{0}$ and the Coulomb coupling $e_c$ in the
Hamiltonian limit.}
\begin{ruledtabular}
\begin{tabular}{ccccc}
 $\lambda$ & $K$ & $e_c$ & $a_{s}/r_{0}$ &  $K$ from ELCE
\protect\cite{Hamer86} \\\hline
 4.2991 & 0.76(3)  & 0.393(4) &0.7790  &0.664(6) \\

 4.6437 & 0.67(2)  & 0.403(4) &0.7354  &0.563(5) \\
 5.0022 & 0.56(1)  & 0.367(3) &0.7120  &0.404(3)\\
 5.1865 & 0.395(9) & 0.324(3) &0.5158  &0.34(2)\\
 5.3741 & 0.265(4) & 0.220(1) &0.4311  &0.24(4)\\
 5.7594 & 0.096(3) & 0.203(1) &0.2580  &0.107(8)\\
 6.1580 & 0.061(2) & 0.245(2) &0.2090  &        \\
 6.5698 & 0.033(2) & 0.262(2) &0.1542  &        \\
 6.9951 & 0.021(1) & 0.251(2) &0.1219  &        \\
\end{tabular}
\end{ruledtabular}
\end{table}

To obtain estimates of glueball masses in the Hamiltonian limit,
an extrapolation of the data points obtained at constant
$\beta_{E}$ for various anisotropies ($\Delta\tau =1 - 0.25$) is
performed in powers of $\Delta\tau^{2}$. Error estimates for the
extrapolation may be obtained by the ``linear, quadratic, cubic"
extrapolation method \cite{Pradeep00}. Figure \ref{figexta1pphm}
shows our estimates for the scalar glueballs as a function of
$\Delta\tau^{2}$ for various fixed $\beta_{E}$ values. Unlike the
string tension, the  scalar mass  depends rather strongly on
$\Delta\tau$. This can be seen from the slope and the curvature in
the extrapolation to the Hamiltonian limit in the plot. This may
be due to the presence of the $O(a_{t}^{2})$ errors in the
standard Wilson action. In the large $\beta_{E}$ region we,
however, see a smooth variation of the scalar mass with
$\Delta\tau^{2}$.
\begin{figure}[!h]
\scalebox{0.45}{\includegraphics{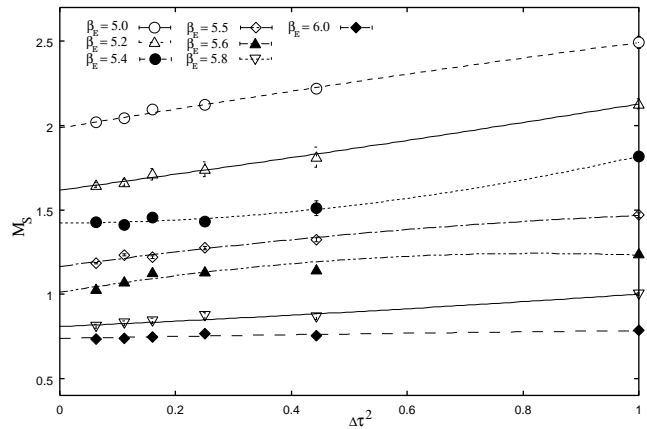}}
\caption{
\label{figexta1pphm}
Extrapolation of the scalar glueball mass (in lattice units)  to the 
Hamiltonian limit
$\Delta\tau\rightarrow 0$, for fixed $\beta_{E}$. The lines show quadratic
fits to the data in $\Delta\tau^{2}$.}
\end{figure}

The extrapolation of axial vector and tensor glueball masses to
the Hamiltonian limit is shown in Figs. \ref{figextt2pphm} and
\ref{figextt1pmhm} respectively.
This is performed again in powers of $\Delta\tau^{2}$, in a
fashion similar to the scalar glueball.
In both these channels we note that both show a
strong dependence on  $\Delta\tau$: in the intermediate $\beta_{E}$ region
there is a factor of 2 - 3 difference between the values at $\Delta\tau =0$
and   $\Delta\tau =1$. Since the leading
discretization errors in the tensor glueball masses are expected to be
$O(a_{t}^{2},a_{s}^{2},\alpha_{s}a_{s}^{2})$, this difference may be
attributed to the presence of $O(a_{t}^{2})$ errors.
However, in the extreme anisotropic region, the results differ very little;
this is obvious since the expected  $O(a_{t}^{2})$ errors decrease as
aspect ratio $\xi=1/\Delta \tau$ is increased.

\begin{figure}[!h]
\scalebox{0.45}{\includegraphics{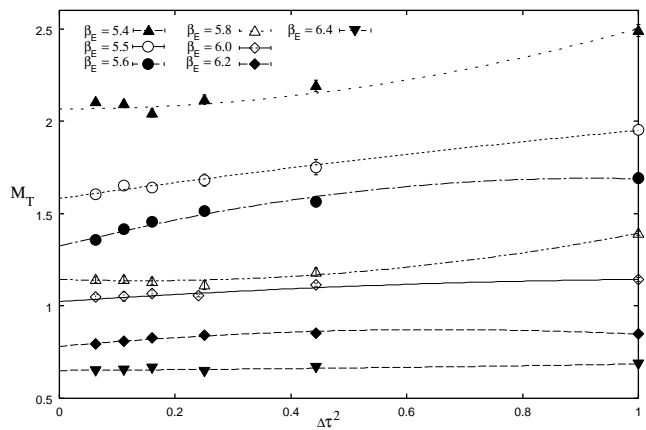}}
\caption{
\label{figextt2pphm}
Extrapolation of the tensor glueball mass (in lattice units) to the 
Hamiltonian limit
$\Delta\tau\rightarrow 0$, for fixed $\beta_{E}$. The lines show quadratic
fits to the data in $\Delta\tau^{2}$.}
\end{figure}

\begin{figure}[!h]
\scalebox{0.45}{\includegraphics{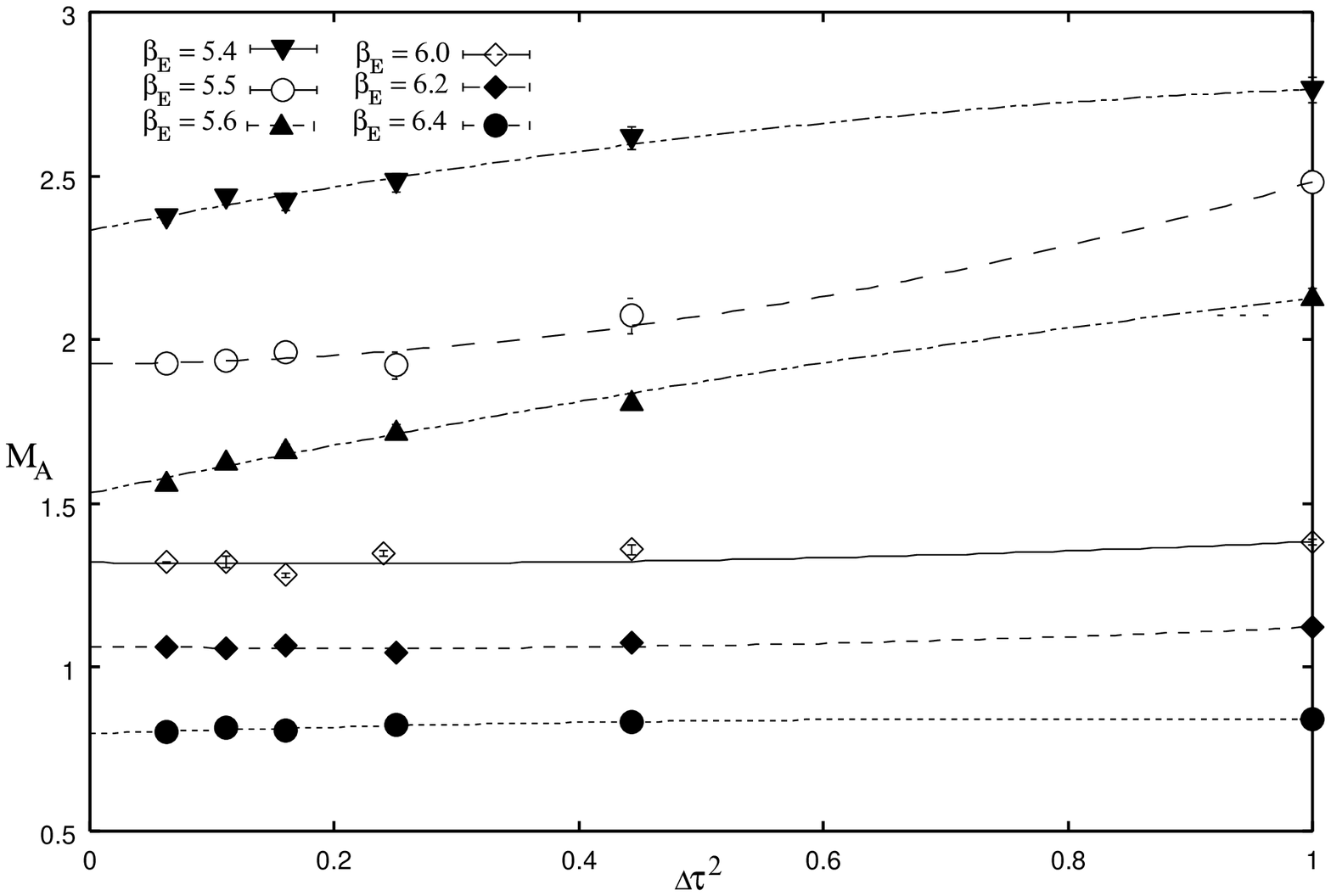}}
\caption{
\label{figextt1pmhm}
Extrapolation of the axial glueball mass (in lattice units) to the 
Hamiltonian limit
$\Delta\tau\rightarrow 0$, for fixed $\beta_{E}$. The lines show quadratic
fits to the data in $\Delta\tau^{2}$.}
\end{figure}

Our Hamiltonian estimates  for the scalar $M_{S}
(=a_{t}m_{0^{++}})$, axial vector $M_{A} (=a_{t}m_{1^{+-}})$ and
tensor $M_{T} (=a_{t}m_{2^{++}})$ glueballs together with the
Hamiltonian coupling $\lambda$, calculated from Eq. (\ref{eqn30a})
are shown in Table \ref{tab_Hmass}.

\begin{table}[!h]
\caption{ \label{tab_Hmass} Glueball masses in the Hamiltonian limit.}
\begin{ruledtabular}
\begin{tabular}{cccc}
$\lambda$  & $M_{S}$ & $M_{A}$ & $M_{T}$ \\\hline
4.2991 & 1.98(1) &  & \\
4.6437 & 1.61(1) &  & \\
5.0022 & 1.42(1) & 2.09(2) & 2.33(2)\\
5.1865 & 1.16(2) & 1.58(1) & 1.80(1)\\
5.3741 & 1.01(2) & 1.40(1) & 1.79(1)\\
5.7594 & 0.84(1) & 1.14(2) & 1.41(2)\\
6.1580 & 0.738(9)& 1.034(9)& 1.32(1)\\
6.5698 & 0.58(1) & 0.78(1) & 1.06(1)\\
6.9951 & 0.47(1) & 0.64(1) &  0.79(1)\\
\end{tabular}
\end{ruledtabular}
\end{table}

\subsection{Asymptotic scaling}
An interesting question concerns the onset of ``scaling'', which
is often interpreted to mean that the physical quantities 
should become independent of the coupling in the
weak-coupling regime. Here, we examine more closely the asymptotic
scaling, given by the two-loop perturbative
$\beta$-function \cite{Osterwnlder78}
\begin{equation}
am = C_m \mbox{exp}\bigg(-\frac{1}{2b_{0}g^{2}}\bigg)\big(
b_{0}g^{2}\big)^{-b_{1}/2b_{0}^{2}}\left[1+O(g^{2})\right],
\label{eqn28}
\end{equation}
where
\begin{displaymath}
b_{0} = \frac{11}{16\pi^{2}}, \hspace{0.50cm}
b_{1}= \frac{102}{(16\pi^{2})^{2}},
\end{displaymath}
with the aim to establish that results approach the expected
scaling form in the weak-coupling regime. The coupling $g$ refers to
the Hamiltonian coupling $g_{H}$ in this case. In general, the most
accurately calculated physical quantity is the string tension $K$,
and hence the dimensionless ratio $m/\sqrt{K}$, since the leading
corrections to such a ratio are known to be of the order
$O(1/\zeta)^{2}$ \cite{Teper98}, where $\zeta$ is some length scale.
However, it is well known that behavior of the string tension is
quite different over the range of coupling: it follows neither
two-loop nor three-loop scaling at these couplings and there is no
universal Callan-Symanzik functions that can accommodate both string
tension and glueball mass down to this coupling range.

Fig. \ref{figscaling} shows the asymptotic scaling behavior of
scalar glueball mass, $M_{S}$, as a function of $\sqrt{\lambda}$. 
The dotted-line is the Hamiltonian strong-coupling
expansion to the order $\beta_{H}^{7}$ \cite{Hamer89} and the
solid line represents a fit to the Hamiltonian weak-coupling
asymptotic form (\ref{eqn28}). It can be seen that our estimates
appear to match nicely onto the strong and weak coupling
expansions in their respective limits. The strong coupling series
is not well behaved as the approximants begin to diverge as the
crossover region is reached, hence allowing only rough estimates.
For comparison we also show the results obtained from plaquette
expansion \cite{Hollenberg98}, which are
found to converge quickly in the strong coupling region and
diverge as the transition region is approached. The results are
somewhat lower than our estimates, but are almost compatible
within errors.

\begin{figure}[!h]
\scalebox{0.45}{\includegraphics{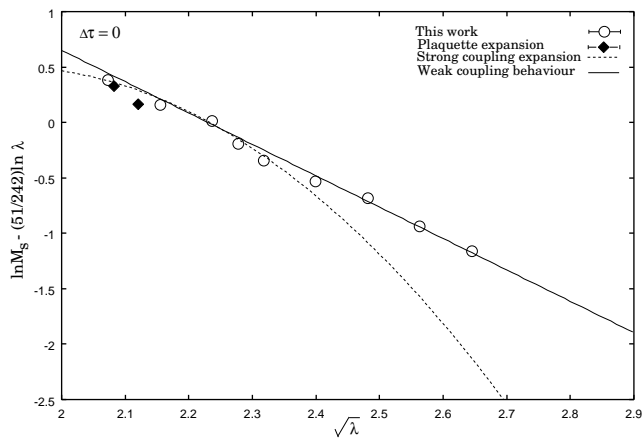}}
\caption{
\label{figscaling}
The asymptotic scaling behavior of scalar mass gap as a function
of Hamiltonian coupling. The errors are smaller than the data points.}
\end{figure}

In the weak-coupling region,  the scalar glueball mass
appears to adhere accurately to the expected asymptotic scaling
behavior.
An unconstrained fit of the form (\ref{eqn28}) to the data
in the range $2.3 \leq \sqrt{\lambda} \leq 2.65$, which corresponds to 
$5.6 \leq \beta_{E} \leq 6.4$, gives a scaling
slope of $2.83\pm 0.17$ which is consistent, within errors,  with  the
predicted value of 2.90.  It is
interesting to note that our estimate of $C_{m} = 260\pm 10$ agrees
well with the value found in plaquette expansion when the curves are matched
onto the scaling form at $\sqrt{\lambda}= 2.02$.

\subsection{Extrapolation to the continuum limit}
To extrapolate the results given
in Table \ref{tab_Hmass} to zero lattice spacing, we combine the glueball mass
estimates with the determinations of the hadronic scale
$r_{0}/a_{s}$. Fig. \ref{figCona1pp} shows the dimensionless product of $r_{0}$
and the scalar glueball mass estimates as a function of
$(a_{s}/r_{0})^{2}$. Also are shown for comparison the earlier Hamiltonian estimates from
the series expansion \cite{Hamer89}, t-expansion \cite{Lana91} and the
plaquette expansion \cite{Hollenberg98}.
\begin{figure}[!h]
\scalebox{0.45}{\includegraphics{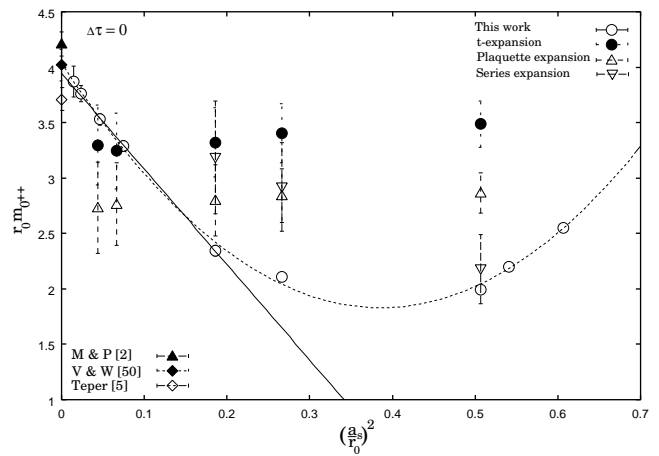}}
\caption{
\label{figCona1pp}
Scalar glueball mass estimates in terms of $r_{0}$ as a function of
the lattice spacing $(a_{s}/r_{0})^{2}$. The solid and dashed lines
represent  linear  and quadratic extrapolation of the data to
the continuum limit.
Results from different methods are also shown for comparison.}
\end{figure}

It can be seen from Fig. \ref{figCona1pp} that scalar glueball mass
shows a peculiar feature. There is no clear sign of the dimensionless
quantity $r_{0}m_{0^{++}}$ reaching to a constant value.
As $a_{s}$ is increased from zero, the scalar mass first decreases,
reaching a minimum near
$a_{s}/r_{0} \approx 0.6$, the mass then gradually increases with $a_{s}$.

The origin of the ``\emph{dip}''
in the scalar glueball mass is attributed to the sensitivity of this state to
discretization errors in the standard Wilson gauge
action. The dip is reduced by about
half when classically improved action \cite{Norman98} is used   and
further reduced when the classical improvement is
supplemented by tadpole improvement \cite{Morningstar97,Morningstar99}.
This implies that the origin
of the dip lies, at least in part, in discretization errors. If
this picture is correct then one might expect the dip to be
further reduced (or even eliminated ) if one accounts for the
perturbative renormalization of action. This would remove the
leading $O(\alpha_{s}a^{2})$ discretization errors in the
tree-level, tadpole-improved action.
However before making a definitive conclusion regarding the observed
sensitivity, one needs to compute the full radiative corrections to the action.
Such calculations are under investigation and we intend to report on these
studies in future.
The fact that the scalar glueball is less sensitive to the finite-volume
effects (Fig. \ref{figfinitevol}) than the tensor state and
somewhat larger lattice spacing
dependence in the scalar mass could indicate that scalar glueball is smaller
in size.

To extract the results in the continuum limit we fit the mass
measurements by the following three-parameter  fitting function:
\begin{equation}
\varphi(a_{s})=r_{0}m_{G}+c_{1}(a_{s}/r_{0})^{2}+c_{2}(a_{s}/r_{0})^{4},
\label{eqnf1}
\end{equation}
where $r_{0}m_{G}$, $c_{1}$ and $c_{2}$ are fit parameters. The
function yields a continuum limit  $r_{0}m_{0^{++}}= 4.03\pm 0.05$
MeV. An extrapolation of the data, in the range $0.014 \leq
(a_{s}/r_{0})^{2} \leq 0.18$ (with $5.6 \leq \beta_{E} \leq 6.4$),
using a linear fit of the form
\begin{equation}
\varphi_{1}(a_{s}) = r_{0}m_{G}+c_{0}(a_{s}/r_{0})^{2}
\label{eqnf2}
\end{equation}
yields $r_{0}m_{0^{++}} = 3.94\pm 0.02$.

A comparison with previous continuum limit estimates obtained by
Morningstar and Peardon \cite{Morningstar99} using tadpole improved action,
Vaccarino and Weingarten \cite{Vaccarino99} and Teper \cite{Teper98} using
standard Wilson action on isotropic lattice  shows that our result
for the scalar mass  is lower
than that obtained by Morningstar and Peardon \cite{Morningstar99} but
in good agreement with result obtained by Vaccarino and Weingarten
\cite{Vaccarino99} and more accurate than the results obtained in
Ref. \cite{Teper98}. We, however,
do not expect a complete agreement with the results obtained in
Ref. \cite{Morningstar99} which were computed for a different action.
A comparison with the earlier Hamiltonian results shows that
our present estimates are higher than those obtained previously.
The $t$-expansion \cite{Lana91} results, obtained from 1/3 D-Pade
approximants, reach a constant value near the continuum and are higher than the
plaquette expansion estimates \cite{Hollenberg98}, obtained by
taking the average of the curves  for various values of set
parameter. A continuum extrapolation of the
$t$-expansion estimates leads to
 $r_{0}m_{0^{++}}= 3.24 \pm 0.025$ whereas the plaquette expansion estimates
give  $2.75\pm 0.018$.  We observe that our result is an excellent
improvement over all previous Hamiltonian results.

\begin{figure}[!h]
\scalebox{0.45}{\includegraphics{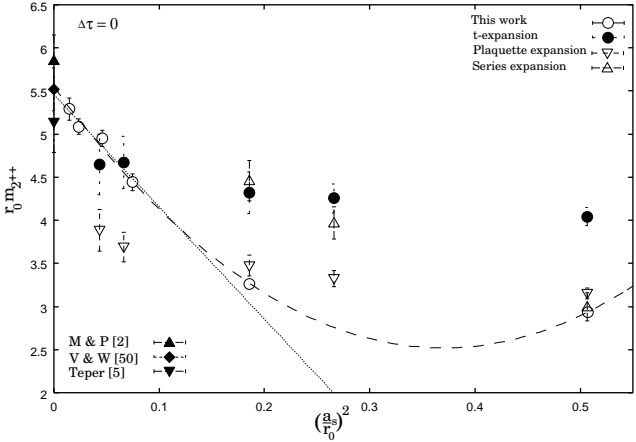}}
\caption{
\label{figConT2}
Same as Fig. {figCona1pp} but for the tensor glueball.}
\end{figure}

Figs. \ref{figConT2} and \ref{figConT1} show, respectively, the continuum limit
extrapolation of the
tensor and axial vector glueball masses in terms of $r_{0}$ against
lattice spacing $(a_{s}/r_{0})^{2}$. Although the leading discretization
errors in the tensor glueball mass are expected to be $O(a_{t}^{2},a_{s}^{2},
\alpha_{s}a_{s}^{2})$, the tensor  glueball data
show less significant finite-spacing errors in contrast to the scalar mass.
This could indicate that the tensor glueball has a larger size.
For our final estimates of the tensor and axial vector glueball masses, we
 perform fits using Eq. (\ref{eqnf1}) while constraining the fit
parameters to be the same for both the fitting functions. The fits yields
$r_0m(2^{++})= 5.54\pm 0.06$ and $r_0m(1^{+-})= 7.17\pm 0.17$ for the
tensor and axial vector glueballs respectively. These results are in agreement
with the estimates obtained in Ref. \cite{Vaccarino99}, however, lower
than the results obtained in Ref. \cite{Morningstar99}. This is
most probably due to the strong curvature in our extrapolations for the tensor
 and axial vector glueballs. In the large $a_{s}$
region our results for the tensor glueballs are in good agreement
with plaquette expansion estimates and  comparable  within errors
to  the $t$-expansion results in the small $a_{s}$ region. Near
the continuum our estimates are higher than previous Hamiltonian
estimates. A linear fits to the $t$-expansion data yield 
continuum values $4.61 \pm 0.08$ and $5.56\pm 0.11$  for $r_{0}m_{2^{++}}$
and $r_{0}m_{1{+-}}$ respectively.

\begin{figure}[!h]
\scalebox{0.45}{\includegraphics{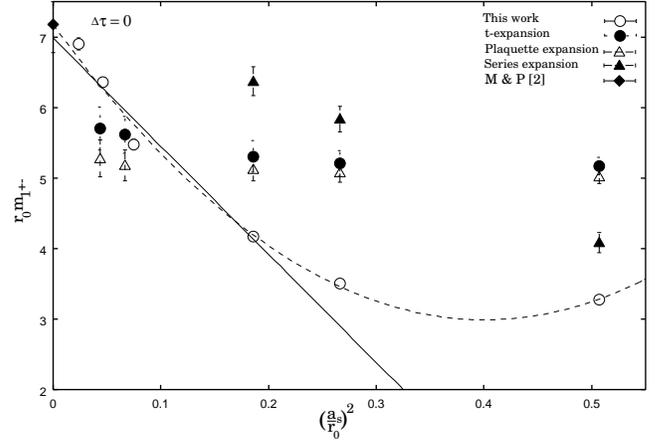}}
\caption{
\label{figConT1}
Same as Fig. {figCona1pp} but for the axial vector glueball mass.}
\end{figure}

The continuum extrapolation of the dimensionless
mass ratio, $R_{T}=m_{2^{++}}/m_{0^{++}}$, which is expected to be
 more easily amenable,  as a function of $(a_{s}/r_{0})^{2}$ is shown in
Figs. \ref{figCT2A0ratio}.
The  mass ratio $R_{T}$ decreases slightly as we move from larger to
 smaller $a_{s}$ and appears
to reach a level near the continuum, but within statistical errors, they can be fitted by a straight limit,
giving $R_{T}= 1.36\pm 0.01$ at $a_s \to 0$. 
In the continuum limit, our
results agree well with the Hamiltonian results
\cite{Hamer89,Lana91,Hollenberg98} .  This is in excellent agreement with
results from the previous studies on Euclidean lattices. Since the
Hamiltonian formulation is thought of as a very asymmetric
Euclidean theory, this agreement provides a test of universality
between the two formulations. It is interesting to note that mass
ratio $R_{T}$ in SU(2) QCD exhibits similar features as that in
SU(3) theory: the SU(2)  $R_{T}$ rises to a value of
approximately 1.5 over the intermediate volume region
\cite{Teper98}. Our estimates
 for the glueball masses and mass  ratio are shown and compared 
with previous results from various other studies  in Table
\ref{tab_finalres}.

\begin{figure}[!h]
\scalebox{0.45}{\includegraphics{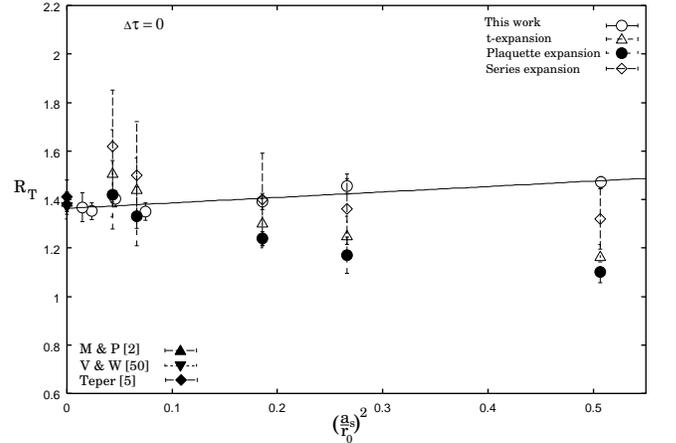}}
\caption{
\label{figCT2A0ratio}
Extrapolation of the mass ratios to the continuum limit. The solid line is a
linear fit to the data.}
\end{figure}

We see the evidence that the axial vector mass is definitely greater than the
scalar and tensor masses for all couplings analyzed here. This is in accordance
with the claim made by Hamer \cite{Hamer89} that  corresponding mass 
ratios are strictly greater than one for all finite couplings.

\begin{table}[!h]
\caption{ \label{tab_finalres}
Glueball masses and the mass ratio in the continuum limit. 
The values indicated
in bold are our final continuum mass estimates. The fitting function to the
data  is given in Eq. (\protect\ref{eqnf1}).}
\begin{ruledtabular}
\begin{tabular}{ccccc}
Reference & $r_{0}m_{0^{++}}$ & $r_{0}m_{2^{++}}$ & $r_{0}m_{1^{+-}}$ & 
 $R_{T}$ \\ \hline
This work & {\bf 4.03(5)}& {\bf 5.54(6)}& {\bf 7.2(2)} & \bf{1.36(2)} \\
M $\&$ P \protect\cite{Morningstar99} &4.21(11) &5.85(2) & 7.18(4)& 1.39(4)\\
V $\&$ W \protect\cite{Vaccarino99} &4.0(1) & 5.5(2) &  &1.37(5)  \\
Teper \protect\cite{Teper98}   &3.65(11) & 5.1(3) &     &1.41(7)   \\
UKQCD \protect\cite{Bali93}    &3.78(12) & 5.53(24) & 6.90(8)& 1.44(6) \\
Hamiltonian \protect\cite{Hamer89,Lana91}  & 3.0 - 3.3 & 4.25 - 4.60 & 
5.40 - 5.80 & 1.2-1.6\\
\end{tabular}
\end{ruledtabular}
\end{table}

Finally, we convert our glueball mass estimates into  physical units by
assigning  value to the hadronic scale. The estimate for $r_{0}$ was obtained
by combining the results from various quenched lattice simulations with
the standard Wilson action \cite{Morningstar97}. The simple average
$r_{0}^{-1} = 410\pm 20$ MeV of the determinations in
Ref. \cite{Morningstar97} was taken as our estimate of the hadronic scale.
For the lowest-lying scalar glueball we
obtain $m(0^{++})= 1654\pm 83 \pm 70$ MeV.
The tensor and axial vector masses, for which the extrapolation to zero
lattice spacing encountered no problems, are
$m(2^{++}) = 2272 \pm 115 \pm 125$ MeV and
$m(1^{+-})=2940 \pm 165\pm 120$ MeV, where
the first error comes from uncertainty in $r_{0}m_{G}$ and second error
comes from the uncertainty in $r_{0}^{-1}$. Our estimates are
in good agreement with those found in Ref. \cite{Vaccarino99} but are
under the predicted values for the resonances $f_{0}(1710)$, $f_{2}(2400)$
by about 10-15$\%$. However, for direct comparison with the experiment, we
have to take into account the contribution from corrections
 due to  the light quark effects and mixing with conventional hadrons:
the quenched approximation will easily lead to an underestimate of
the simulated values. A possible solution to this problem is to construct a
QCD Hamiltonian that couples conventional hadrons and glueballs of identical
$J^{PC}$  so that physical states are a linear combination of hadron and
glueball states. This will shed some light on the contents and
dominant components of the predicted resonances in the scalar and excited
glueball states. The glueball with exotic quantum numbers will, however,
not mix with the conventional hadrons and would be ideal for establishing
the existence of glueballs.

\section{CONCLUSIONS}
\label{secVI}

In this work we used standard Euclidean MC methods 
to extract the Hamiltonian limit of
4-dimensional SU(3) LGT.
By taking the renormalization of both the anisotropy and couplings
we have calculated the string tension and glueball masses in the
extreme anisotropic limit. We have found that the renormalization
of the couplings and anisotropy has a large influence on
extrapolating our results to the Hamiltonian limit
$\Delta\tau\rightarrow 0$. It has been suggested that this
renormalization has a much smaller effect for improved actions
\cite{Morningstar97,Morningstar99}, however later calculations
\cite{Sakai00} showed that the discrepancy between the couplings
$\beta_{\xi}$ and $\beta_{E}$ reach a maximum near high
anisotropies ($\Delta\tau \leq 0.333$). It would be interesting to
see how this renormalization influences the extrapolation of the
results for improved actions to the Hamiltonian limit.

Estimates for the scalar, tensor and axial vector glueballs were
extrapolated to the continuum limit and the results are presented
in terms of the hadronic scale $r_{0}^{-1}$. Extrapolation of the
scalar glueball mass to the continuum limit was problematic than
those of tensor and axial vector masses. In the continuum limit 
the mass ratio $R_{T}$  
was observed to scale to $1.36\pm 0.2$. 
We have demonstrated that there is a broad agreement between our results 
and the results
obtained in the Euclidean limit of the theory and a substantial
improvement over previously known estimates calculated within the
Hamiltonian formulation. This demonstrates clear evidence of the
universality between the Euclidean and Hamiltonian formulations.
Our main aim for examining anisotropic lattice was to investigate
alternative MC procedures for obtaining reliable results
in the Hamiltonian limit, in view of the lack of a reliable
numerical method for Hamiltonian LGT. We have
found that standard Euclidean MC approach is more
successful than other quantum MC methods, in particular
the Greens Function MC techniques. 

From the accuracy of
the results obtained here we conclude that standard MC
approach is a preferred technique for calculating physical
observables in the Hamiltonian limit, just as in the Euclidean
formulation. In order to make the Euclidean MC method a
more valuable tool in the Hamiltonian LGT, it will be crucial to
show that it allows one to treat eventually matter fields in the
non-Abelian case. The major disadvantage of this approach is the
cost in computer time, as several configuration sets must be
generated to obtain one data point.

\begin{acknowledgments}

This work was done on our PC cluster as
well as the LSSC2 cluster. We thank E. Gregory, C.
Hamer, and C. Morningstar for a number of valuable suggestions
which provided the impetus for much of this work.
We are also grateful to T. Xiang for useful discussions. This work was
supported by the Key Project of National Natural Science
Foundation (10235040), Key Project of Chinese Ministry of
Education (105135), and Guangdong Provincial Ministry of Education.
\end{acknowledgments}


\end{document}